\documentclass[fleqn,usenatbib,usedcolumn]{mnras}

\usepackage{newtxtext,newtxmath}
\usepackage{xfrac}

\usepackage[T1]{fontenc}

\usepackage{color, soul}

\DeclareRobustCommand{\VAN}[3]{#2}
\let\VANthebibliography\thebibliography
\def\thebibliography{\DeclareRobustCommand{\VAN}[3]{##3}\VANthebibliography}

\usepackage{graphicx}	
\usepackage{amsmath}	
\usepackage[normalem]{ulem}

\usepackage{xcolor}
\usepackage{hyperref}
\newcommand{\orc}{\includegraphics[height=\fontcharht\font`A]{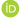}}
\newcommand{\orcid}[1]{\href{https://orcid.org/#1}{\orc}}

\newcommand{\rebound}{{\tt rebound}}
\newcommand{\astropy}{{\tt astropy}}
\newcommand{\numpy}{{\tt numpy}}
\newcommand{\matplotlib}{{\tt matplotlib}}
\newcommand{\scipy}{{\tt scipy}}
\newcommand{\kepler}{{\it Kepler}}

\newcommand{\abin}{a_{\mathrm{bin}}}
\newcommand{\ebin}{e_{\mathrm{bin}}}
\newcommand{\ibin}{i_{\mathrm{bin}}}
\newcommand{\Obin}{\Omega_{\mathrm{bin}}}
\newcommand{\ip}{i_P}
\newcommand{\Op}{\Omega_P}

\DeclareRobustCommand{\VAN}[3]{#3}

\title[Transiting misaligned circumbinary planets]{
The Number of Transits Per Epoch \\for Transiting Misaligned Circumbinary Planets
}

\author[Chen \& Kipping]{
Zirui Chen\orcid{0000-0001-8755-3836}$^{1}$\thanks{E-mail: zc2445@columbia.edu (ZC)} \&
David Kipping\orcid{0000-0002-4365-7366}$^{1}$
\\
$^{1}$Dept. of Astronomy, Columbia University, New York, NY 10027, USA\\
}

\date{Accepted XXX. Received YYY; in original form ZZZ}

\pubyear{2022}

\begin{document}
\label{firstpage}
\pagerange{\pageref{firstpage}--\pageref{lastpage}}
\maketitle

\begin{abstract}
The growing catalog of circumbinary planets strengthens the notion that planets form in a diverse range of conditions across the cosmos. Transiting circumbinary planets yield especially important insights and many examples are now known, in broadly coplanar obits with respect to their binary. Studies of circumbinary disks suggest misaligned transiting examples could also plausibly exist, but their existence would exacerbate the already challenging feat of automatic detection. In this work, we synthesize populations of such planets and consider the number of transits per epoch they produce, forming integer sequences. For isotropic distributions, such sequences will appear foreign to conventional expectation, rarely (${\sim}1$\%) producing the signature double-transits we’ve come to expect for circumbinaries, instead producing sparse sequences dominated by zero-transit epochs (${\sim}80$\%). Despite their strangeness, we demonstrate that these sequences will be non-random and that the two preceding epochs predict the next to high accuracy. Additionally, we show that even when clustering the transits into grouped epochs, they often appear unphysical if erroneously assuming a single star, due to the missing epochs. Crucially, missing epochs mean highly isotropic populations can trick the observer into assigning the wrong period in up to a quarter of cases, adding further confusion. Finally, we show that the transit sequences encode the inclination distribution and demonstrate a simple inference method that successfully matches the injected truth. Our work highlights how the simple act of flagging transits can be used to provide an initial, vetting-level analysis of misaligned transiting circumbinary planets.
\end{abstract}

\begin{keywords}
binaries: general -- techniques: photometric -- planet and satellites: detection
\end{keywords}



\section{Introduction}
\label{sec:intro}

Binary star systems represent a manifestly frequent outcome of star formation
\citep{pringle:1989,chapman:1992,tohline:2002}. Indeed, the majority of
Solar type stars reside within binary or higher multiplicity systems
\citep{karttunen:1994,raghavan:2010} - defying preconceptions one might foster
through arguments of mediocrity. This naturally raises questions about the
prevalance of planets around such binaries, which may exist in so-called
``S-type'' (orbiting just one of the stars) or ``P-type'' (orbiting both)
configurations \citep{dvorak:1986}. For binary stars separated by less than
50\,AU, observations reveal a clear dearth of S-type planets
\citep{kraus:2016}, although counter-examples such as $\gamma$ Cephei Ab
\citep{hatzes:2003} and KOI-1257 Ab \citep{santerne:2014} demonstrate that a
population does indeed exist. Further, simulation work concurs that planet
formation is expected to be generally suppressed for S-type orbits in close
binaries (e.g. see \citealt{thebault:2015} and references therein).

For P-type orbits, so-called circumbinary planets (or even ``Tatooines''),
the situation is quite different. Many circumbinary planets have been
discovered via transits \citep{welsh:2012}, characterised by their elaborate
light curves featuring one or more dips per epoch, of varying depths,
durations and timings (e.g. \citealt{doyle:2011}). Unlike S-type planets,
occurrence rate estimates suggest that giant circumbinary planets
($>6$\,$R_{\oplus}$) orbiting within 300\,days are just as common as single
stars \citep{armstrong:2014}.

Two key observations from this population are highlighted. The first is that
there is a conspicuous absence of planets around the tightest main sequence
binaries, consistent with such binaries forming through a phase of destructive
inward migration \citep{munoz:2015,hamers:2016,xu:2016}. The second is that
there is a dearth of short-period circumbinary planets, and a pile-up
close to a critical period of dynamical instability \citep{welsh:2014}.
This boundary of instabiltiy was characterised with a simple semi-empirical
formula in \citet{holman:1998} but was later extended to account for resonance
instabilities independently by \citet{lam:2018} and \citet{quarles:2018}.

Despite these advances, our view of these systems may not be complete. In particular, there
may be a missing population of circumbinary planets - those which are
misaligned to their binary. \citet{armstrong:2014} note that their occurrence
rate estimate should be treated as a lower-limit, since a misaligned population
could not be ruled out. \citet{martin:2014} show that misaligned systems would
lead to even stranger transit sequences, featuring epochs with missing transits
that would be challenging to identify in the first place. At first glance, such
transit sequences may ostensibly appear ``random'', making automated discovery
difficult. The ``Random Transiter'' (HD 139139) exemplifies\footnote{We note that
here both P- and S-type planetary transits were indeed hypothesised by
\citet{rappaport:2019}, but ultimately found to inadequately explain their
data.} the challenge of working with irregular sequences like this
\citep{rappaport:2019}. So difficult these sequences are that
\citet{martin:2018} comments that ``transit discovery methods that are
sensitive to highly misaligned planets are yet to be demonstrated''. 

A population of inclined circumbinary planets would thus clearly prove a
challenge to existing detection algorithms, but would also provide unique
insights into these enigmatic systems. Imaging of young binary systems provides
some guidance as to whether such a population might exist, by measuring the
alignment between the protoplanetary disks (from which planets form) and the
binary (measured spectrosopically/astrometrically). \citet{czekala:2019}
constrain that 68\% of circumbinary disks are aligned to within $3^{\circ}$
for binaries with periods shorter than 20\,days, but beyond 30\,days one
finds a diverse range up to polar orbiting disks. Whilst our knowledge here
is clearly incomplete, a population of inclined circumbinary planets is
consistent with current observations.

Conventional transit search algorithms, such as the ``Box Least Squares'' (BLS)
algorithm \citep{kovacs:2002}, utterly fail for even aligned circumbinary
planets. This is because they assume strict periodicity between transits, but
the binary's constantly changing position imparts large timing and duration
changes. As a result, early \kepler\ detections were largely conducted by
eye \citep{doyle:2011,welsh:2012}. A modification of BLS, designed for planets
orbiting single stars plus some timing perturbation, dubbed QATS
(``Quasiperiodic Automated Transit Search'') offers a possible improvement, but
still does not account for the variable durations or the possibility of two
dips \citep{carter:2013}. Efforts to modify BLS specifically for circumbinary
planets have previously been made, such as the method of \citet{klagyivik:2017}
that allows both the depth and duration of the transits to freely vary.
However, the use of a large timing window to scan over introduces a higher
false positive rate. \citet{windemuth:2019} offer the most advanced model to
date, dubbed QATS-EB, using a simplified, semi-analytic model for the binary
motion. However, even here, the method is forced to make numerous simplifying
assumptions which may limit the sensitivity to extreme systems.

Clearly, as noted by \citet{windemuth:2019}, one could relax these limitations
by conducting searches using a fully photodynamical model, but the high
dimensionality of the problem makes such a search impractical. Simplified
models are clearly needed to enable the rapid analysis of ever growing
databases of light curves, and in that context we might begin by simply
asking - \textit{what do the light curves of transiting misaligned
circumbinary planets look like?}

Perhaps the most basic observable one might consider in this effort is the
number of transits that occur each epoch. For a single star this always
equals unity. But, as noted earlier, aligned circumbinary planets generally
cause two transits \citet{martin:2014}. During a syzygy event, one transit would
be expected and for a very short-period binary and long-period planet one could
observe more than two transits. Misaligned systems add even more complexity though by
permitting zero-transit epochs. This makes automated detection immediately
more challenging since now the transit sequence becomes sparse, and creates
``broken'' sequences where even the clustered transits don't appear to follow
a linear ephemeris. Indeed, such sequences might appear ostensibly random
and be rejected by conventional algorithms demanding some pattern in the data.

In this work, then, we tackle the issue as to how many transits per epoch
one expects for transiting misaligned circumbinary planets. If the transits
are individually of sufficient high signal-to-noise, it should be possible
to construct a catalog of the transit times which could then be clustered into
groups representative of epochs. Individual transits are demonstrably
discoverable either by citizen scientists (e.g. \citet{eisner:2021}) or an
automated routine sensitive to single events of potentially highly peculiar
shape (such as the ``weird detector'' algorithm; \citealt{wheeler:2019}).
This work seeks to build some intuition and understanding of how these
sequences would appear in our data, and how they could be ultimately used
to make progress in the detection and interpretation of a misaligned
circumbinary population.

In Section~\ref{sec:nonrandom}, we demonstrate that non-isotropic circumbinary
planet populations are non-random by developing a memory slot prediction
algorithm and showing that it outperforms random predictions. In
Section~\ref{sec:transitcounts}, we establish a relationship between the
observed transits per epoch counts and the inclination distribution of a
circumbinary planet population. Based on this relationship, we develop a
Bayesian framework to infer the inclination distribution of a circumbinary
planet population. In Section~\ref{sec:discussion}, we conclude by discussing the challenges and opportunities that misaligned circumbinary systems pose to future works.

\section{Randomness of Circumbinary Planet Transit Sequences}
\label{sec:nonrandom}

\subsection{Random Worlds}

\begin{figure}
\includegraphics[width=\columnwidth]{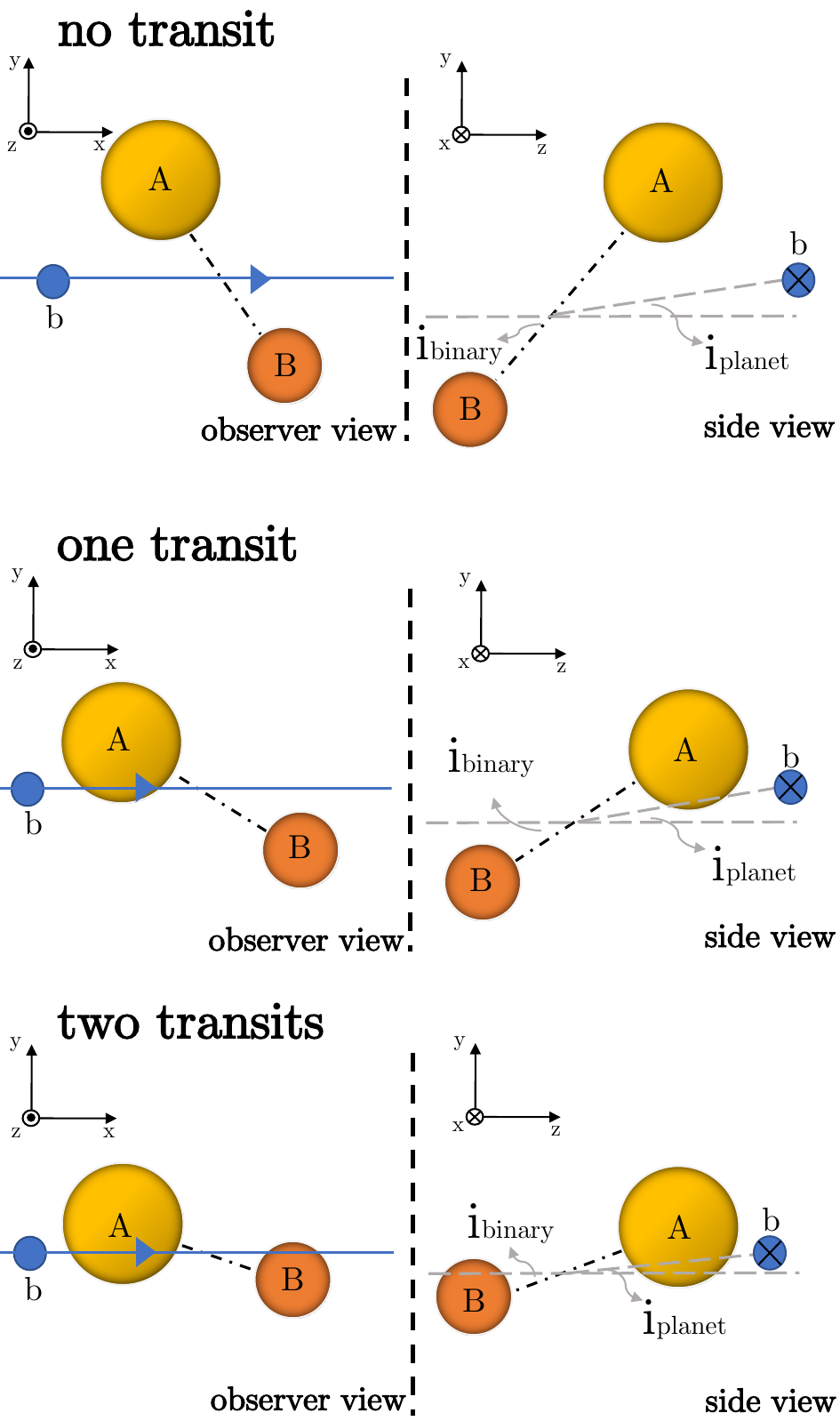}
\caption{Schematic diagram showing how different configurations of misaligned
circumbinary planets can lead to different transit counts per epoch. ``A''
and ``B'' represent the two stars, and ``b'' represents the circumbinary planet.
This diagram is not to scale and only for illustrative purposes.}
\label{fig:setup}
\end{figure}

As discussed in Section~\ref{sec:intro}, this paper primarily concerns - how
many transits per epoch does one expect for transiting misaligned circumbinary
planets? The number of transits per epoch is perhaps the simplest observable
one might reasonably define for such systems, yet we lack a good understanding
of its behaviour. Figure~\ref{fig:setup} graphically illustrates how a
circumbinary planet can cause zero, one or two transits. Indeed, more than two transits
is also possible if the binary motion is sufficiently fast to ``catch up'' with
the planet again during a given epoch. This case is challenging to graphically
illustrate, and thus is not shown explicitly in Figure~\ref{fig:setup}, but it
is fully considered in what follows.

In light of this, a series of transit epochs will manifest a potentially
variable number of transits each time. For example, the first epoch might
yield two transits, the next epoch yields zero, the third epoch just a
single dip, and so on. Such a series could be compactly represented using the
digits ``201...''. From the observer's perspective, the true nature of the
system is of course not immediately obvious; they are simply faced with a
particular sequence and need to try to interpret the underlying cause.

The ``Random Transiter'' \citep{rappaport:2019} again provides an analogous
example of this situation. Rather than seeing a sequence of clustered
events of varying counts (as is the case for circumbinary planets),
\citet{rappaport:2019} instead report non-clustered transits scattered across
the time series. Despite this difference, the problem is ostensibly similar to
ours in that the lack of any clear pattern in the transits was confounding
and unusual. A basic conclusion of their work is that the sequence
was consistent with being produced from a random number generator. If indeed
the transit times are truly stochastic, then it would be impossible to
ever predict when the next transit would occur based on the previous
data. This determination of randomness, if verified with future data, is
important because one might hypothesise that complex, chaotic systems (e.g.
stellar activity) to feasibly produce (outwardly) stochastic events like this.
Thus, the determination of whether a given series of astrophysical observations
are random or not is a crucial lens through which we can make progress in
interpreting such data.

To this end, we here begin by asking if for circumbinary planets the transit
counts per epoch, represented as a sequence of integers, yields a sequence that
would (at least outwardly) appear random or not? To address this, we develop
and test a memory slot prediction algorithm (see Section~\ref{sec:memoryalg})
and show that it outperforms random predictions for non-isotropic circumbinary
planet populations. This establishes that non-isotropic circumbinary planet
populations are non-random.

\subsection{Population Synthesis}
\label{sec:popgen}

In order to investigate the randomness (or lack thereof) for circumbinary
transit sequences, we first need a large, correctly-labelled population of
such systems to study. Due to the limited number of detected circumbinary
planets, and the need for accurate labels, we generated a population of
artificial circumbinary planet systems by specifying parameters of the
binary and the planetary orbit. In what follows, we describe how we
generated each system.

We first randomly pick an observed \kepler\ target from the Data Release 25 (DR25)
Kepler Input Catalog queried using the Mikulski Archive for Space Telescopes
(MAST). The queried stellar mass is adopted as the true mass of the primary
star in our circumbinary system. In this way, the mass of our primary stars are
representative of the mass distribution of the \kepler\ sample.

Next, the mass ratio, $q$, of the binary stars is assumed to follow a $q^{0.5}$
distribution following \citep{parker:2013}. After assigning $q$, we are then
able to trivially determine the mass of the second star. The stellar radii (which have little effect on our results) can
then be approximated using the empirical mass-radius relation of main sequence
stars given by \citet{demircan:1991}:

\begin{equation}
R =
\begin{cases}
1.06 M^{0.945} & \text{for M < 1.66 $M_{\odot}$}\\
1.33 M^{0.555} & \text{for M > 1.66 $M_{\odot}$}\\
\end{cases}       
\end{equation}

In what follows, it assumed that the gravitational influence of the planet on
the binary is negligible and thus the planet is treated as a massless particle
of one Earth radius. Further, it is assumed that the eccentricity of both the
binary and the planetary orbit is zero.

The binary separation is randomly drawn from a log-uniform distribution between
a predefined minimum and maximum value. As a starting point, the minimum binary
separation among all the observed circumbinary systems is 0.08\,AU and we adopt
a minimum of 0.05\,AU in what follows. For the maximum, we are only interested
in stable planets around binary stars. Thus, the maximum binary separation is
defined to be the separation such that a planet just outside the instability
region defined by \citet{holman:1998} has a period of one \kepler\ window
(4.35\,years). With a larger binary separation, even the innermost stable
circumbinary planet cannot be detected by \kepler\ because its period is too
long.

For the semi-major axis of the planet, we again set a minimum and maximum
possible value and sample in between following a
log-uniform distribution. The minimum semi-major axis of a stable planetary
orbit is given by \citet{holman:1998} as:

\begin{align}
&a_P \geq \abin (1.60 + 5.10\ebin - 2.22\ebin^2 + 4.12\mu \nonumber \\
& - 4.27\ebin\mu -  5.09\mu^2 + 4.61 \ebin^2\mu^2)
\label{eq:quadratic}
\end{align}

where $\abin$ is the semi-major axis of the binary orbit, $\ebin$ is the
eccentricity of the binary orbit (in our case fixed to zero), and
$\mu = \frac{m_B}{m_A + m_B}$, where $m_A$ is the mass of the more massive
star, and $m_B$ is the mass of the less massive star. On the other hand, the
maximum semi-major axis is defined such that the corresponding planetary
period is equal to one \kepler\ window. Beyond this maximum semi-major axis,
the planet would be unobservable.

The binary star's orbit should be distributed isotropically (i.e. no preferred orientation) because there is nothing special about the observer's line of sight. To generate such an isotropically distributed population of binary star orbits, we take caution in specifying the binary orbit's longitude of ascending node, $\Obin$, and inclination
angle, $\ibin$ (defined with respect to the observer's line of sight).

First, consider the inclination angle. An isotropic population might naively appear to follow a uniform distribution, but this would in fact produce vectors non-uniformly distributed across the celestial sphere (due to the curved geometry). Instead, an isotropic distribution is truly described by a uniform distribution in $\cos \ibin$. On the other hand, the longitude of ascending node of the binary orbit, $\Obin$, is sampled uniformly to yield an isotropic distribution. To ensure that we cover all possible orbital configurations, we uniformly sample $\cos \ibin$ between -1 and 1 and uniformly sample $\Obin$ between $-\frac{\pi}{2}$ and $\frac{\pi}{2}$.

With the binary orbit's longitude of ascending node, $\Obin$, and inclination angle, $\ibin$, sampled as prescribed above, we now considered these two parameters for the planetary orbit - $\ip$ and $\Op$.
First consider $\ip$. The inclination of the planetary orbit with respect to the binary orbit (rather than the observer's line of sight) characterise the coplanarity of the system, which is intimately related to the expected transit sequences. For example, if a population of circumbinary planets were perfectly coplanar with their host binary's orbit, one would expect to get transits over every epoch if the system is favorably aligned in an "edge-on" orbit with respect to the observer. If, on the other hand, a population of circumbinary planets were perfectly isotropic (i.e. no preferred orientation) with respect to their host binary's orbit, then one would expect transits to be less frequent and more erratic in nature. Next, consider $\Op$. Whether or not $\Op$ is related to coplanarity as well depends on how it's defined. If we define $\Op$ with respect to the binary orbit, which means we always rotate the planetary orbit around the axis perpendicular to the binary orbital plane by $\Op$, then $\Op$ is unrelated to coplanarity. That is, the orbits are coplanar for any value of $\Op$ as long as $\ip$ and $\ibin$ are identical. However, If we define $\Op$ with respect to the observer's line of sight, then the difference between $\Op$ and $\Obin$ is closely related to coplanarity, where the case with $\Op=\Obin$ corresponds to a coplanar orbit. To avoid confusion and for simpler implementation, we choose the second definition, which is to define $\Op$ with respect to the observer's line of sight. Finally, we note that the "coplanarity" being to referred to here is a characteristic of the
\textit{population} of circumbinary planets, not of an individual planet. To
generate such a population, we require a flexible, analytic probability
distribution describing $\Op$ and $\ip$, from which one can sample.

As mentioned in the earlier discussion on $\ibin$, an isotropic distribution is truly described by a uniform distribution in the cosine of the inclination angle. The same holds true if we want to produce an isotropic population of planets. However, as discussed before, it is the relative inclination angle that characterises coplanarity. For simplicity of notation, we denote the relative inclination angle as $\Delta i$, where $\Delta i = \ip - \ibin$. A coplanar
population is described by a highly compressed (approaching a Dirac Delta
function) distribution centred on $\mathrm{E}[\Delta i] = 0$, or equivalently, $\mathrm{E}[\frac{\pi}{2} - \Delta i ] = \frac{\pi}{2}$ and 
$\mathrm{E}[\cos \big(\frac{\pi}{2} - \Delta i\big)] = 0$. On the other hand, we take a uniform distribution in the cosine of $\frac{\pi}{2} - \Delta i$ for an isotropic distribution, where we use the angle $\frac{\pi}{2} - \Delta i$ for consistency. Since a truncated normal has the
flexibility to produce highly compressed, diffuse and even quasi-uniform
distributions, we adopted this form in what follows. Accordingly, one may
write that $\cos \big(\frac{\pi}{2} - \Delta i\big) \sim \bar{\mathcal{N}}[0,\sigma_{\cos \big(\frac{\pi}{2} - \Delta i\big)}]$, where
$\bar{\mathcal{N}}$ denotes a truncated normal distribution (with support from
$-1$ to $+1$ and centered at $0$) and $\sigma_{\cos \big(\frac{\pi}{2} - \Delta i\big)}$ is the standard
deviation, such that $\sigma_{\cos \big(\frac{\pi}{2} - \Delta i\big)} \to 0$ is consistent with a coplanar
population and $\sigma_{\cos \big(\frac{\pi}{2} - \Delta i\big)} \to \infty$ is consistent with an isotropic
population. Finally, to convert between $\Delta i$ and $\ip$ (inclination of the planetary orbit measure with respect to the observer's line of sight), we simply use the relationship $\ip = \ibin + \Delta i$, where $\ibin$ is obtained from an isotropic distribution as discussed above.

For $\Op$, it is again crucial to realize that it is the relative longitude of ascending node that characterises coplanarity. We denote the relative longitude of ascending node as $\Delta \Omega$ (where $\Delta \Omega = \Op - \Obin$) and elect to also distribute the longitude of the ascending
node of the planet following another truncated Gaussian. We limit ourselves to consider only prograde planets (planets orbiting in the same direction as the binary stars), which means we want to sample $\Delta \Omega$ between $-\frac{\pi}{2}$ and $\frac{\pi}{2}$ instead of the complete $-\pi$ to $\pi$ range. To do so, we sample from a truncated Gaussian
such that $\Delta \Omega/(\frac{\pi}{2}) \sim \bar{\mathcal{N}}[0,\sigma_{\Delta \Omega}]$ (with support from $-1$ to
$+1$ and centered at $0$), where now by virtue of normalizing
by $\frac{\pi}{2}$ we maintain the same truncation limits as with $\cos \big(\frac{\pi}{2} - \Delta i\big)$. Finally, we can convert between $\Delta \Omega$ and $\Op$ using the relationship  $\Op = \Obin + \Delta \Omega$, where $\Obin$ is obtained from an isotropic distribution as discussed above.

These two parameters, $\cos \big(\frac{\pi}{2} - \Delta i\big)$ and $\Delta \Omega$, both describe the
degree of non-coplanarity, but along different dimensions. For example, a
coplanar population would correspond to $\sigma_{\\cos \big(\frac{\pi}{2} - \Delta i\big)} \to 0$ and
$\sigma_{\Delta \Omega} \to 0$. In contrast, an isotropic population corresponds to
$\sigma_{\cos \big(\frac{\pi}{2} - \Delta i\big)} \to \infty$ and $\sigma_{\Delta \Omega} \to \infty$. 
In this way, and because of the carefully chosen parameterisation that
maintains the same truncation limits, the degree of coplanarity can be
encapsulated with just a single term, $\sigma \equiv \sigma_{\cos \big(\frac{\pi}{2} - \Delta i\big)} =
\sigma_{\Delta \Omega}$. To provide some context, the $\sigma$ of all
\textit{observed} transiting circumbinary planets is $\sim 5 \times 10^{-4}$
(noting that $\sigma$ is in fact a dimensionless quantity).

\begin{figure}
\includegraphics[width=\columnwidth]{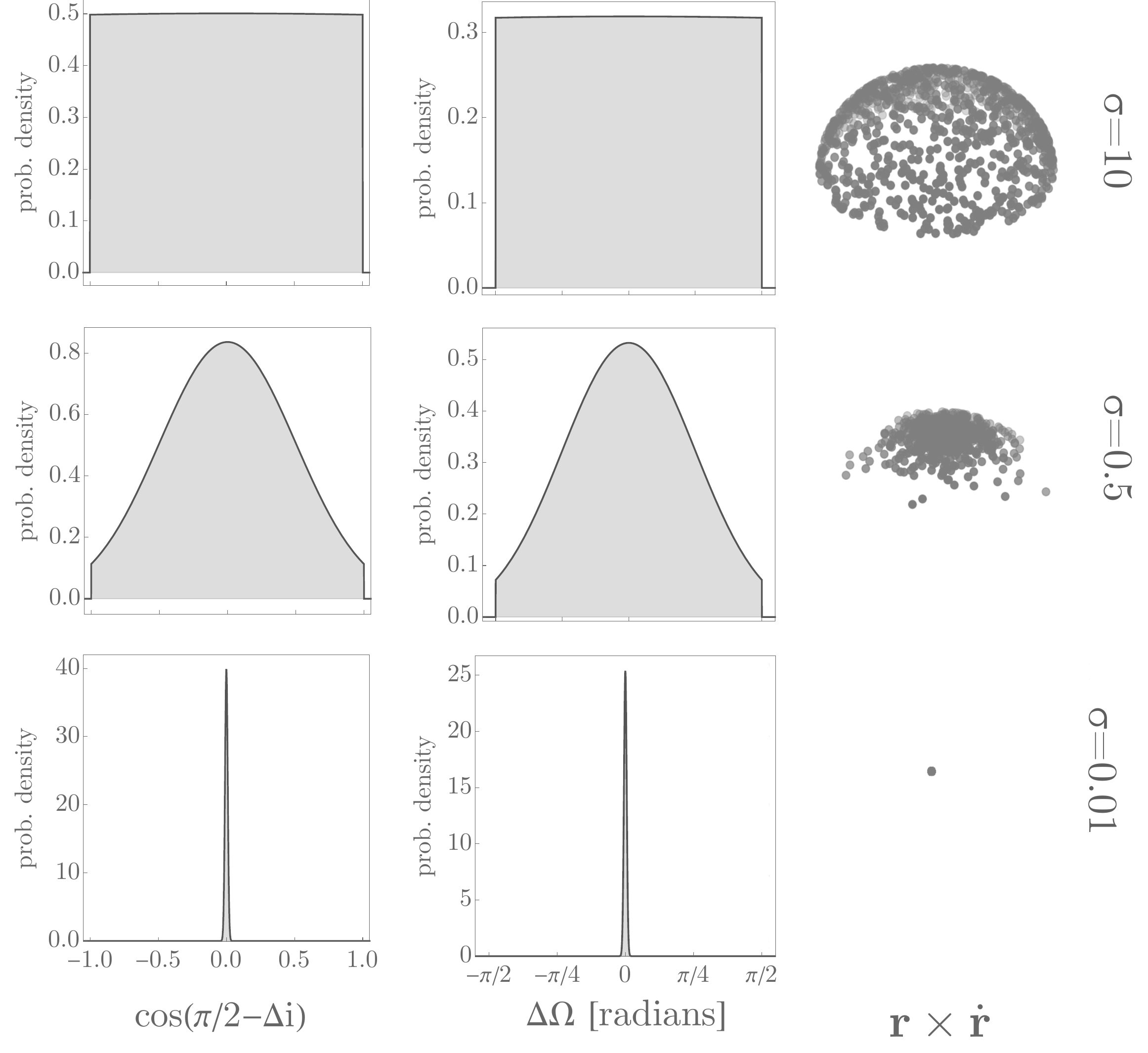}
\caption{
Behaviour of the cosine of the relative inclination (left) and the relative longitude of the ascending node
(middle) probability distributions as a function of $\sigma$, for three choices
(three rows). In the right column, we visualise fair samples for the vector
$\mathbf{r}\times\dot{\mathbf{r}}$ of the planetary orbit, fixing the binary orbit on the equatorial plane of the celestial sphere for easier visualisation.}
\label{fig:histogram}
\end{figure}

In Figure~\ref{fig:histogram}, we consider three representative values of
$\sigma$: $\sigma = 10$ (corresponding to an isotropic population), $\sigma =
0.3$ (corresponding to an intermediate population between coplanar and
isotropic), and $\sigma = 10^{-2}$ (corresponding to a coplanar population).
For each value of $\sigma$, we show the corresponding probability distributions
for $\cos \big(\frac{\pi}{2} - \Delta i\big)$ and $\Delta \Omega$. One can see that these, as well as the
distribution of normal vectors on the globes, are consistent with the expected
behaviours in limiting cases.

Now that all the relevant orbital parameters are established, we make the additional note that these parameters only set the initial configuration of a circumbinary planet system. They are by no means constant and can evolve over time according to the dynamics of the system. This time-dependence of the orbital parameters is fully captured by our simulations.

We are now ready to generate a random circumbinary system. Because we
are investigating the effect of changing $\sigma$, these systems are of
course not representative of the underlying distribution, whose $\sigma$
is not well-characterised. But the stellar masses and binary separations
are designed to be quasi-representative (especially once we filter out
unphysical systems in the next subsection).

We use the \rebound\ package \citep{rein:2012} to evolve the circumbinary
system over a time span much longer than one \kepler\ observing window and
record the number of transits over each epoch. Note that we do not
distinguish between the two stars when counting transits. In practice, it is
often possible to distinguish the identity of the star being transited -
especially for $q \ll 1$ since the depths become increasingly distinct (e.g.
\citealt{doyle:2011}). However, the misaligned nature of our systems means that
the impact parameters are also constantly changing, leading to V-shaped
transits and variable depths even for U-shaped events due to
limb darkening \citep{mandel:2002}. This complicates the process of flagging
identities to each event and introduces the possibility of mis-labelled events.
In contrast, we argue that the simple act of counting how many dips are present
in the data is far less ambiguous and thus presents a more robust labelling
system, at the expense of some compromise in information content. Of course,
even here there is the potential for some degree of confusion due to
overlapping transits, but broadly one should expect the number of dips to be
interpretable in such cases (e.g. see \citealt{gillon:2017}).

\subsection{Filtering the Synthesised Sequences}
\label{sec:filtering}

After generating and evolving each of the circumbinary systems, we are left
with a sequence of integers that denotes the number of transits per epoch for
each planet. Next, we impose several filters to ensure that we are left with
systems that are both physically sensible and observable.

Among all the observed circumbinary planets, the shortest period is
${\simeq}50$\,days (Kepler-47b; \citealt{orosz:2012}). Accordingly, we impose
this as a constraint in our artificial circumbinary population and filter out
all planets with periods shorter than 50\,days.

We can only extract information from circumbinary planets that are observable
to us. In what follows, an ``observable'' planet is defined to be one that
undergoes at least two epochs with at least one transit in one \kepler\
observing window. Thus, we need to filter out all unobservable planets.

To determine if a planet is ``observable'' or not, we first calculate the
number of epochs it undertakes during one \kepler\ window and denote it as $n$.
Then, we randomly extract a sub-sequence of length $n$ from the full transits
per epoch sequence we obtained from our simulation. If this sub-sequence does
not contain at least two non-zero entries, them we discard it because it would
outwardly lack the two necessary to cross our threshold of being
``observable''. We repeat this procedure for all possible sub-sequences with
length $n$ and filter out all the unobservable sub-sequences. 

For an isotropic circumbinary planet population, $\sim8$\% of the systems passes through the filters mentioned above and contain at least one length $n$ sub-sequence that is ``observable''. On the other hand, only $\sim0.4$\% of the systems in a coplanar planet population contain observable sub-sequences. The fact that coplanar systems are less likely to be observable seems counter-intuitive, but we note that for a coplanar system to be observable, the inclination of its binary and planet orbit must be precisely aligned to the observer's line of sight. On the other hand, since the inclination of the binary and planet orbit is not correlated in the isotropic case, the requirement for observable systems is less stringent.

\subsection{A Prediction Algorithm Using Memory}
\label{sec:memoryalg}

Recall that the underlying question tackled in this section is - do the
transits per epoch sequences of transiting misaligned circumbinary planets
follow a pattern which is discernible from being purely random? The
hypothesis of a stochastic sequence could be undermined if were possible
to make predictions for the next number in a sequence. Even if the accuracy of said predictions are not
perfect, the demonstrable ability to predict better than random would still
establish that the sequence was not truly stochastic in nature. To this
end, we introduce a simple algorithm for predicting the
number of transits a circumbinary planet undergoes in a given epoch using
information from the previous epochs (hence the name "memory slot“), and show that the predictive power of this algorithm is
consistently better than random guesses, thus proving that the transit
occurrences of circumbinary planets are non-random and tractable.

\subsubsection{Training and Testing the Algorithm}

We are working with a sequence of integers that denotes the number of transits
per epoch for the planet, and we wish to be able to predict the value of any of
the integers in the sequence.

Taking inspiration from context networks (e.g. \citealt{sandford:2021}), we
postulate that there is memory in this sequence. In particular, we will use a
two memory slot prediction algorithm (see Section~\ref{sec:doublememory} for
further justification). That is, we use the $(n-2)^{\mathrm{th}}$ and
$(n-1)^{\mathrm{th}}$ integer in the sequence to predict the $n^{\mathrm{th}}$
integer.

\begin{figure*}
\includegraphics[width=550pt]{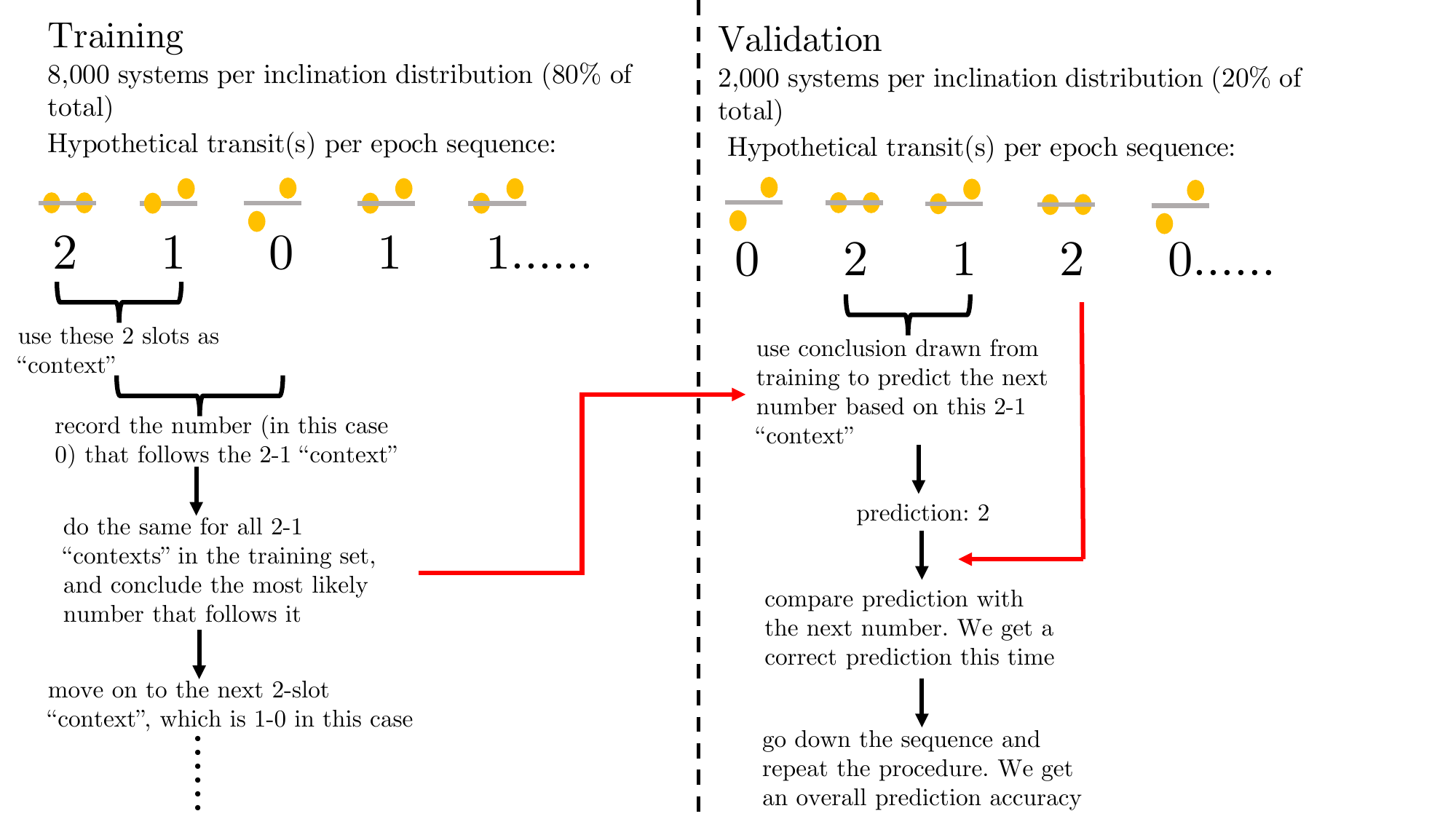}
\caption{Illustrative graphic depicting the process which by our two-slot
memory algorithm is trained and validated.}
\label{fig:training}
\end{figure*}

To ``learn'' how the predictive model works, we divide the dataset of
transits per epoch sequences for a population of circumbinary planets into two
parts: 80\% goes into a training set, and 20\% goes into a validation set. To
predict the $n^{th}$ integer in a sequence in the validation set, we first
identify the two-integer sequence consisting of the $(n-2)^{th}$ and
$(n-1)^{th}$ integer that precedes the prediction slot. Then, we look for this
two-integer sequence in the training set. For each occurrence of this sequence,
we record the integer that immediately follows. Finally, we predict the
$n^{th}$ integer to be the integer that has the highest probability (i.e. modal
outcome) of following the $(n-2)^{th}$ and $(n-1)^{th}$ integer sequence in the
training set. We direct the reader to Figure~\ref{fig:training} for a
graphic decomposition of the procedure described above.

\subsubsection{Why Two Memory Slots?}
\label{sec:doublememory}

In our prediction algorithm, we are using the two preceding slots to inform our
prediction. We have also tried using more or fewer slots. With one memory slot,
the prediction power of our algorithm was found to modestly decrease, with accuracy dropping by up to 7\% depending on the assumed $\sigma$ value.
With more than two memory slots, the prediction power is almost identical to
that of using only two memory slots, with accuracy increasing by up to 3\% depending on the assumed $\sigma$ value.

This indicates that the ``memory'' of circumbinary transits is primarily
contained in the two preceding slots. Further, it demonstrates that amongst the
two preceding slots, the one that is immediately before the slot of interest
contains a larger fraction of the ``memory''. Thus, using a two memory slot
prediction algorithm ensures that we obtain the maximum prediction power with
the least effort.

\subsection{Comparing the Performance of the Prediction Algorithm}
\label{sec:benchmark}

With the predictive model defined, we next turn to a point of comparison.
We consider two possible "benchmark" models to this end, which we introduce
in what follows. Further, we introduce the important concept of "broken
sequences".

It is straightforward to apply the memory slot prediction algorithm described
in Section~\ref{sec:memoryalg} to the validation set and measure the
corresponding prediction accuracy. However, this prediction accuracy in isolation means
very little since it remains unclear whether our model establishes
non-randomness to the sequences. To better understand and contextualise this
prediction accuracy, it is useful to compare it to a benchmark
accuracy, derived using a "naive" algorithm (following
\citealt{sandford:2021}).

We considered two possible benchmarks. The first is obtained using a simplistic
algorithm: predict the mode of the dataset every time. In this way, the model
simply predicts the same guess every time once ``trained''. As a second model,
we try taking the observed sequence and simply drawing a random choice from it
as the ``prediction'' for the next epoch. We dub these benchmark models as the
``modal'' and ``random'' models. Our prediction algorithm can claim to
have predictive power only if it can out-perform these benchmarks. 

\subsection{Broken (and Unbroken) Sequences}
\label{sec:broken}

In considering the types of sequences that might be observed, one concern is
that many sequences might be discarded by automated surveys as seemingly
impossible. This is particularly poignant for misaligned circumbinary
systems, since the binary star may not be eclipsing and thus its binary nature
may not even be known beforehand. A sequence such as ``1111'' would appear
clearly reasonable, even with large timing variations, as a likely
single planet transiting a single star with some additional hidden perturber
(e.g. see \citealt{nesvorny:2012}). A sequence such as ``1121'' would also
be quite reasonable, with the ``2'' event possibly caused by a second
longer period transiting planet or even an exomoon (e.g.
\citealt{teachey:2018}) - which we note need not transit each epoch
\citep{martin:2019x}. However, a sequence such as ``1101'' is far
more challenging to understand and apparently without precedent (although
the lack of precedence may simply be a product of the fact that such sequences
are unexpected and thus possibly rejected). With the ``1121'' case,
we can cluster the transits within each epoch together and still draw a
quasi linear ephemeris through the times of the clustered events,
consistent with the exomoon hypothesis for example. But a ``1101''
sequence produces a sparse chain that cannot be represented by even
a quasi linear ephemeris. We consider here that such sequences,
which we dub ``broken sequences'', may trip over many automated searches
and could plausibly be absent in future catalogs.

We define ``broken sequences'' as those for which the time gap between
observable epochs (i.e. non-zero numbers of transits) is not constant
(and thus an ``unbroken sequence'' is the opposite case).

For example, a transits per epoch sequence of ``1011'' or ``100101''
corresponds to a broken system, and a sequence of ``1211'' or ``10201''
corresponds to an unbroken system. The fraction of simulations that
generate broken sequences is thus an important parameter when considering
the occurrence rate and completeness estimates, in cases where
this feature would indeed negatively impact the detectability of such
sequences in surveys.

We find that the ratio of broken to unbroken sequences is strongly
affected by the coplanarity of the circumbinary population generated. Varying
the governing $\sigma$ parameter in Figure~\ref{fig:brokens}, we find a
log-logistic type behaviour emerges. This reveals that for quasi-isotropic
populations, some 30\% of sequences may present broken sequences in \kepler-like
observing windows. Accordingly, surveys designed to measure the occurrence
rate and properties of transiting circumbinary planets need to either use
algorithms capable of identifying broken sequences or correct their rates
for this effect appropriately.

As noted, the fraction of broken systems versus $\log(\sigma)$ curve in
Figure~\ref{fig:brokens} resembles a logistic function. This can be understood
by the fact that at extreme values of $\sigma$, the distribution saturates
to either perfectly isotropic or perfectly coplanar.

\subsection{Aliased (and Unaliased) Sequences}
\label{sec:aliased}

Broken systems could potentially cause a survey to reject a system as
unphysical, although this of course depends upon the search algorithms being
utilised. A sequence such as ``1101'' could be plausibly reconstructed though
and the correct period inferred. However, the broken sequence ``1010001''
would be challenging to assign the true period for without detailed modelling
or additional information; in particular we only observe the 1's and so thus
the simplest assumption is that the underlying sequence is ``1101''. In such
a case, we would have assigned an orbital period which is twice as long as the
true one - an aliased period. Of course, this aliasing issue can also arise on
unbroken sequences, such as the sequence ``10101''. Thus, when interpreting
such systems, besides from dividing the sequences up between broken and
unbroken, we can also delineate them between ``aliased'' and ``unaliased''.

This aliasing is important to our predictive model because it effectively
transforms sequences like ``1010001'' $\to$ ``1101'', which will clearly
influence the training and validation of our proposed algorithm. However,
if the true period were known, perhaps through detailed modeling or
additional observations such as radial velocity measurements, then this
transformation would not occur and we could use our algorithm without
modification.

\begin{figure}
\includegraphics[width=\columnwidth]{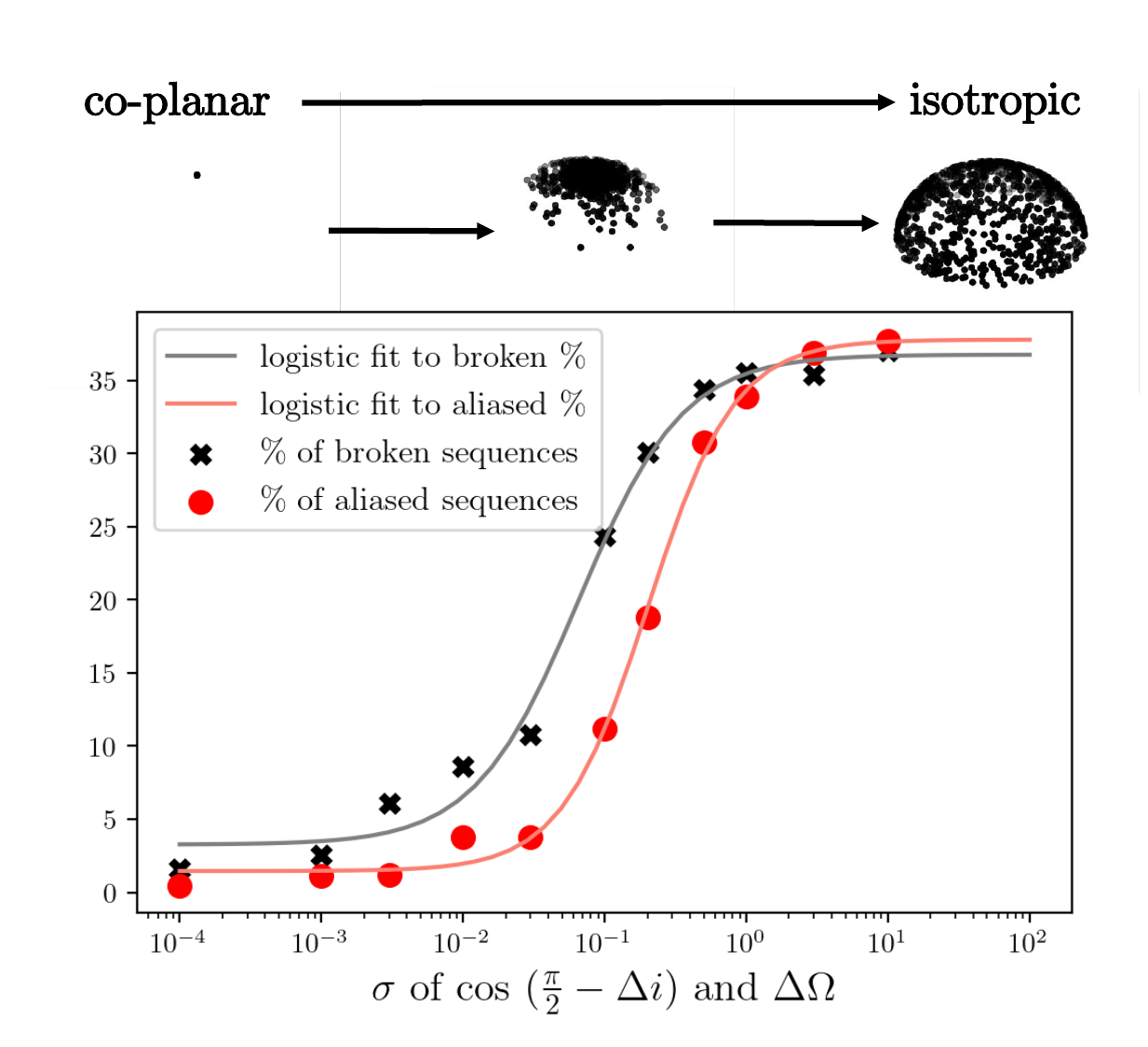}
\caption{
The fraction of so-called broken and aliased sequences recorded amongst eleven simulated populations of transiting circumbinary planets, each with differing degrees of coplanarity to the binary star (characterised by the $\sigma$ parameter). For each $\sigma$, 10,000 circumbinary systems are generated and analysed. Each dot on the globes represents a
$\mathbf{r}\times\dot{\mathbf{r}}$ vector of the planetary orbit, fixing the binary orbit on the equatorial plane of the celestial sphere for easier visualisation. These vectors (and thus planetary orbital planes) on the three globes are sampled according to different inclination distributions, ranging from co-planar to isotropic.
}
\label{fig:brokens}
\end{figure}

As with broken sequences, the ratio of aliased and unaliased sequences exhibits
a log-logistic type behaviour as we vary the governing $\sigma$ parameter in
Figure~\ref{fig:brokens}. For quasi-isotropic populations, more than a quarter
of the sequences are aliased. This not only affects the training
and validation of our proposed algorithm, but also implies that extra caution
needs to be applied when assigning a period to observed transiting circumbinary
planets.

\subsection{Results}
\label{sec:results}

To summarise, we have proposed a simple predictive model for the
number of transits per epoch based on the previous two epochs. In
addition, we have proposed two benchmark predictive models that
simply exploit the ensemble statistics rather than the last two
epochs, one that always returns the mode and one that chooses a random
sample from the list. In the results that follow, we always compare
the performance of our full predictive model to these simpler two
as a benchmark.

Further, we note that observed sequences can be delineated in two
different ways. First, a sequence is ``broken'' if a quasi linear
ephemeris cannot explained even the clustered sequence of transits
(otherwise its unbroken). Such cases could potentially be missed
by detection algorithms and thus deserve special consideration.
Second, a sequence could lead one to assign an aliased period for
the planet rather than the true one. In this way systems can
be broken or unbroken, and also aliased or unaliased.

In evaluating the performance of our predictive model against the
benchmark models, there are thus a total of 4 different scenarios
that could be considered

\begin{itemize}
\item[{\textbf{A]}}]
Broken+unbroken sequences, where the true period is correctly assigned.
\item[{\textbf{B]}}]
Broken+unbroken sequences, where the apparent period is assigned (which may be
aliased).
\item[{\textbf{C]}}]
Unbroken sequences only, where the true period is correctly assigned.
\item[{\textbf{D]}}]
Unbroken sequences only, where the apparent period is assigned (which may be
aliased).
\end{itemize}

For each of these four scenarios, we start from the transits per epoch sequences
obtained from generating and evolving populations of circumbinary planets as
prescribed in Section~\ref{sec:popgen}. Then we filter out the broken sequences
and transform the aliased sequences to reflect the apparent period when
necessary. The code we used to treat each of the four scenarios is made publicly
available at this
\href{https://github.com/ziruichen11/misaligned_circumbinary_planets}{URL}.

For each of the four scenarios, we compare the performance of our predictive
model with the two benchmark models described in Section~\ref{sec:benchmark}.
Our results are summarised in Tables~\ref{tab:scenarioA}, ~\ref{tab:scenarioB},
~\ref{tab:scenarioC}, and ~\ref{tab:scenarioD}.

\begin{table*}
\caption{
Summary of simulation results and prediction algorithm performances, here shown
for Scenario A (broken+unbroken sequences and true period is correctly
assigned). Each row shows a different choice for $\sigma$ (left-panel), which
characterises the coplanarity of the system (small $\sigma$ is more coplanar).
Middle-panel shows the frequency of each possible outcome for the number of
transits per epoch, across our suite of simulations. Right-panel shows the
performance of three predictive algorithms for the number of transits per epoch.
}
\centering 
\begin{tabular}{c c c c c c c c c c c} 
\hline\hline 
$\sigma$ & \vline &
\vtop{\hbox{\strut frequency}\hbox{\strut of 0/epoch}} &
\vtop{\hbox{\strut frequency}\hbox{\strut of 1/epoch}} &
\vtop{\hbox{\strut frequency}\hbox{\strut of 2/epoch}} &
\vtop{\hbox{\strut frequency}\hbox{\strut of 3/epoch}} &
\vtop{\hbox{\strut frequency}\hbox{\strut of 4+/epoch}} & \vline &
\vtop{\hbox{\strut two memory}\hbox{\strut slot accuracy}} &
\vtop{\hbox{\strut modal benchmark}\hbox{\strut \,\,\,\,\,\,\,\,\,\,accuracy}} &
\vtop{\hbox{\strut random benchmark}\hbox{\strut \,\,\,\,\,\,\,\,\,\,accuracy}} \\ [0.5ex]
\hline 
		10 & \vline & 80.4\% & 18.6\% & 0.953\% & 0.030\% & 0\% & \vline & 81.3\% & 80.4\% ($\hat{n}=$0) & 64.4\%\\
		3 & \vline & 79.8\% & 19.1\% & 1.10\% & 0.063\% & 0\% & \vline & 80.7\% & 79.8\% ($\hat{n}=$0) & 63.3\%\\
		1 & \vline & 79.5\% & 19.4\% & 1.08\% & 0.014\% & 0\% & \vline & 80.4\% & 79.5\% ($\hat{n}=$0) & 62.2\%\\
		0.5 & \vline & 78.3\% & 19.7\% & 1.82\% & 0.126\% & 0\% & \vline & 80.7\% & 78.3\% ($\hat{n}=$0) & 61.5\%\\
		0.2 & \vline & 73.5\% & 22.8\% & 3.14\% & 0.565\% & 0.079\% & \vline & 78.7\% & 73.5\% ($\hat{n}=$0) & 54.4\%\\
		0.1 & \vline & 66.5\% & 26.9\% & 5.75\% & 0.626\% & 0.179\% & \vline & 78.2\% & 66.5\% ($\hat{n}=$0) & 45.9\%\\
		0.03 & \vline & 43.6\% & 36.3\% & 18.1\% & 1.37\% & 0.562\% & \vline & 77.5\% & 43.6\% ($\hat{n}=$0) & 35.3\%\\
		0.01 & \vline & 26.8\% & 41.4\% & 26.1\% & 4.28\% & 1.38\% & \vline & 81.9\% & 41.4\% ($\hat{n}=$1) & 31.7\%\\
		0.003 & \vline & 9.25\% & 53.2\% & 32.6\% & 2.95\% & 2.01\% & \vline & 87.2\% & 53.2\% ($\hat{n}=$1) & 39.4\%\\
        0.001 & \vline & 2.46\% & 55.1\% & 36.7\% & 4.21\% & 2.24\% & \vline & 89.7\% & 55.1\% ($\hat{n}=$1) & 43.9\%\\
        0.0001 & \vline & 1.70\% & 44.7\% & 44.5\% & 5.09\% & 3.25\% & \vline & 90.5\% & 44.7\% ($\hat{n}=$1) & 40.0\%\\ [1ex]
\hline\hline 
\end{tabular}
\label{tab:scenarioA} 
\end{table*}

\begin{table*}
\caption{Same as Table~\ref{tab:scenarioA} except for Scenario B
(broken+unbroken sequences and the apparent period is assigned, which may be aliased).
} 
\centering 
\begin{tabular}{c c c c c c c c c c c} 
\hline\hline 
$\sigma$ & \vline &
\vtop{\hbox{\strut frequency}\hbox{\strut of 0/epoch}} &
\vtop{\hbox{\strut frequency}\hbox{\strut of 1/epoch}} &
\vtop{\hbox{\strut frequency}\hbox{\strut of 2/epoch}} &
\vtop{\hbox{\strut frequency}\hbox{\strut of 3/epoch}} &
\vtop{\hbox{\strut frequency}\hbox{\strut of 4+/epoch}} & \vline &
\vtop{\hbox{\strut two memory}\hbox{\strut slot accuracy}} &
\vtop{\hbox{\strut modal benchmark}\hbox{\strut \,\,\,\,\,\,\,\,\,\,accuracy}} &
\vtop{\hbox{\strut random benchmark}\hbox{\strut \,\,\,\,\,\,\,\,\,\,accuracy}} \\ [0.5ex]
\hline 
		10 & \vline & 72.1\% & 26.4\% & 1.41\% & 0.017\% & 0\% & \vline & 77.7\% & 72.1\% ($\hat{n}=$0) & 49.8\%\\
		3 & \vline & 72.0\% & 26.9\% & 1.46\% & 0.180\% & 0\% & \vline & 77.8\% & 72.0\% ($\hat{n}=$0) & 49.6\%\\
		1 & \vline & 71.5\% & 26.7\% & 1.78\% & 0.004\% & 0\% & \vline & 77.6\% & 71.5\% ($\hat{n}=$0) & 50.5\%\\
		0.5 & \vline & 72.3\% & 25.4\% & 2.18\% & 0.150\% & 0.013\% & \vline & 78.9\% & 72.3\% ($\hat{n}=$0) & 51.6\%\\
		0.2 & \vline & 69.5\% & 26.5\% & 3.53\% & 0.479\% & 0.050\% & \vline & 79.5\% & 69.5\% ($\hat{n}=$0) & 48.4\%\\
		0.1 & \vline & 64.4\% & 28.4\% & 6.41\% & 0.593\% & 0.241\% & \vline & 79.1\% & 64.4\% ($\hat{n}=$0) & 43.7\%\\
		0.03 & \vline & 46.0\% & 34.9\% & 17.4\% & 1.31\% & 0.432\% & \vline & 79.2\% & 46.0\% ($\hat{n}=$0) & 34.0\%\\
		0.01 & \vline & 23.5\% & 43.4\% & 26.7\% & 4.54\% & 1.83\% & \vline & 83.2\% & 43.4\% ($\hat{n}=$1) & 31.9\%\\
		0.003 & \vline & 8.32\% & 51.3\% & 35.1\% & 2.96\% & 2.32\% & \vline & 89.6\% & 51.3\% ($\hat{n}=$1) & 38.9\%\\
        0.001 & \vline & 2.06\% & 56.8\% & 35.4\% & 3.72\% & 2.04\% & \vline & 91.4\% & 56.8\% ($\hat{n}=$1) & 45.9\%\\
        0.0001 & \vline & 1.58\% & 42.7\% & 46.8\% & 5.36\% & 3.56\% & \vline & 91.8\% & 46.8\% ($\hat{n}=$2) & 41.1\%\\ [1ex]
\hline\hline 
\end{tabular}
\label{tab:scenarioB} 
\end{table*}

\begin{table*}
\caption{Same as Table~\ref{tab:scenarioA} except for Scenario C
(unbroken sequences and the true period is correctly assigned).
} 
\centering 
\begin{tabular}{c c c c c c c c c c c} 
\hline\hline 
$\sigma$ & \vline &
\vtop{\hbox{\strut frequency}\hbox{\strut of 0/epoch}} &
\vtop{\hbox{\strut frequency}\hbox{\strut of 1/epoch}} &
\vtop{\hbox{\strut frequency}\hbox{\strut of 2/epoch}} &
\vtop{\hbox{\strut frequency}\hbox{\strut of 3/epoch}} &
\vtop{\hbox{\strut frequency}\hbox{\strut of 4+/epoch}} & \vline &
\vtop{\hbox{\strut two memory}\hbox{\strut slot accuracy}} &
\vtop{\hbox{\strut modal benchmark}\hbox{\strut \,\,\,\,\,\,\,\,\,\,accuracy}} &
\vtop{\hbox{\strut random benchmark}\hbox{\strut \,\,\,\,\,\,\,\,\,\,accuracy}} \\ [0.5ex]
\hline 
		10 & \vline & 83.0\% & 15.7\% & 1.25\% & 0.057\% & 0\% & \vline & 84.8\% & 83.0\% ($\hat{n}=$0) & 66.1\%\\
		3 & \vline & 82.4\% & 16.5\% & 0.943\% & 0.119\% & 0\% & \vline & 84.2\% & 82.4\% ($\hat{n}=$0) & 64.7\%\\
		1 & \vline & 82.3\% & 16.6\% & 1.07\% & 0.011\% & 0\% & \vline & 84.2\% & 82.3\% ($\hat{n}=$0) & 64.8\%\\
		0.5 & \vline & 82.8\% & 15.9\% & 1.23\% & 0.087\% & 0\% & \vline & 85.0\% & 82.8\% ($\hat{n}=$0) & 64.4\%\\
		0.2 & \vline & 75.7\% & 20.7\% & 3.10\% & 0.513\% & 0.048\% & \vline & 82.8\% & 75.7\% ($\hat{n}=$0) & 54.7\%\\
		0.1 & \vline & 66.9\% & 25.3\% & 6.78\% & 0.895\% & 0.158\% & \vline & 81.4\% & 66.9\% ($\hat{n}=$0) & 45.1\%\\
		0.03 & \vline & 45.3\% & 35.5\% & 17.4\% & 1.32\% & 0.537\% & \vline & 80.9\% & 45.3\% ($\hat{n}=$0) & 34.0\%\\
		0.01 & \vline & 24.9\% & 39.6\% & 26.4\% & 6.70\% & 2.32\% & \vline & 85.9\% & 39.6\% ($\hat{n}=$1) & 30.9\%\\
		0.003 & \vline & 5.32\% & 54.2\% & 35.2\% & 3.19\% & 2.13\% & \vline & 91.9\% & 54.2\% ($\hat{n}=$1) & 42.0\%\\
        0.001 & \vline & 1.60\% & 52.6\% & 40.0\% & 4.88\% & 2.38\% & \vline & 91.6\% & 52.6\% ($\hat{n}=$1) & 43.6\%\\
        0.0001 & \vline & 0.141\% & 43.1\% & 47.5\% & 4.87\% & 3.41\% & \vline & 92.6\% & 47.5\% ($\hat{n}=$2) & 40.3\%\\ [1ex]
\hline\hline 
\end{tabular}
\label{tab:scenarioC} 
\end{table*}

\begin{table*}
\caption{Same as Table~\ref{tab:scenarioA} except for Scenario D
(unbroken sequences and the apparent period is assigned, which may be aliased). We note that in this scenario, broken sequences are removed and
the remaining unbroken sequences undergo an aliasing transform. This means that there are no zeros in the final sequences.
} 
\centering 
\begin{tabular}{c c c c c c c c c c c} 
\hline\hline 
$\sigma$ & \vline &
\vtop{\hbox{\strut frequency}\hbox{\strut of 0/epoch}} &
\vtop{\hbox{\strut frequency}\hbox{\strut of 1/epoch}} &
\vtop{\hbox{\strut frequency}\hbox{\strut of 2/epoch}} &
\vtop{\hbox{\strut frequency}\hbox{\strut of 3/epoch}} &
\vtop{\hbox{\strut frequency}\hbox{\strut of 4+/epoch}} & \vline &
\vtop{\hbox{\strut two memory}\hbox{\strut slot accuracy}} &
\vtop{\hbox{\strut modal benchmark}\hbox{\strut \,\,\,\,\,\,\,\,\,\,accuracy}} &
\vtop{\hbox{\strut random benchmark}\hbox{\strut \,\,\,\,\,\,\,\,\,\,accuracy}} \\ [0.5ex]
\hline 
		10 & \vline & 0\% & 94.1\% & 5.89\% & 0\% & 0.033\% & \vline & 83.8\% & 94.1\% ($\hat{n}=$1) & 90.0\%\\
		3 & \vline & 0\% & 95.0\% & 4.52\% & 0.439\% & 0\% & \vline & 87.2\% & 95.0\% ($\hat{n}=$1) & 91.0\%\\
		1 & \vline & 0\% & 94.3\% & 5.69\% & 0\% & 0\% & \vline & 91.2\% & 94.3\% ($\hat{n}=$1) & 90.5\%\\
		0.5 & \vline & 0\% & 91.9\% & 7.42\% & 0.634\% & 0\% & \vline & 85.0\% & 91.9\% ($\hat{n}=$1) & 88.2\%\\
		0.2 & \vline & 0\% & 86.3\% & 11.7\% & 1.74\% & 0.225\% & \vline & 81.4\% & 86.3\% ($\hat{n}=$1) & 78.9\%\\
		0.1 & \vline & 0\% & 77.5\% & 19.8\% & 2.62\% & 0.119\% & \vline & 78.3\% & 77.5\% ($\hat{n}=$1) & 67.4\%\\
		0.03 & \vline & 0\% & 63.6\% & 33.2\% & 2.31\% & 0.838\% & \vline & 75.3\% & 63.6\% ($\hat{n}=$1) & 55.3\%\\
		0.01 & \vline & 0\% & 59.6\% & 35.2\% & 2.91\% & 2.26\% & \vline & 87.6\% & 54.6\% ($\hat{n}=$1) & 44.6\%\\
		0.003 & \vline & 0\% & 55.8\% & 38.9\% & 2.88\% & 2.43\% & \vline & 94.7\% & 55.8\% ($\hat{n}=$1) & 45.0\%\\
        0.001 & \vline & 0\% & 56.2\% & 36.4\% & 4.46\% & 2.92\% & \vline & 92.6\% & 57.2\% ($\hat{n}=$1) & 47.1\%\\
        0.0001 & \vline & 0\% & 43.9\% & 47.6\% & 5.23\% & 3.30\% & \vline & 91.7\% & 47.6\% ($\hat{n}=$2) & 41.0\%\\ [1ex]
\hline\hline 
\end{tabular}
\label{tab:scenarioD} 
\end{table*}

\subsection{Interpreting the results}

The results from this exercise reveal several important insights. First, the
``random'' version of the benchmark algorithm performs worse than the ``modal'' version for isotropic distributions, but the accuracies of the two benchmarks converge as we approach the coplanar distributions.  For isotropic cases, the large of number of zeros clearly favours the modal method but as the sequences become more varied, the two methods converge in accuracy.

However, as a second key observation, we note that our two memory slot
predictions almost always out-perform either benchmark. The only exception is
in scenario D for the highly isotropic cases, where the modal case wins out.
We note that the number of one's is exceptionally high in this regime, reaching
94\%, and thus the modal algorithm achieves excellent accuracy purely as a result
of the monotonous nature of the sequences. For near coplanar cases, the two
memory slot method reaches an accuracy of ${\sim}92$\% where as the benchmark methods cannot do better than ${\sim}50$\%.

As a result of this second observation, we can arrive at a conclusion concerning
the original motivation behind this section. The number of transits per epoch
for misaligned (or aligned) circumbinary transiting planets does not follow a
purely random sequence and can be effectively predicted. As before, we emphasise
that our prediction algorithm is not optimal and encourage work to develop
improved iterations.

\section{Transit Number Distribution vs. Inclination}
\label{sec:transitcounts}

We have seen how our $\sigma$ parameter, which describes the degree of
coplanarity of a population of systems, has a strong influence on the transit
sequences that result. Of course, this analysis comes from a position of
population synthesis where we know the true $\sigma$ value (because we chose
it). This raises the question as to how the inverse operation might proceed, to
take an observed population and infer $\sigma$. 

Already we have seen that there is a relationship between $\sigma$ and the
distribution of transits per epoch counts, reflected through the results shown
in Section~\ref{sec:results}. This provides a pathway to potentially inferring
the underlying alignment distribution by examining the distribution of transit
counts. 

Intuitively, this makes sense if we consider the extreme cases: for a perfectly coplanar circumbinary population, the observable planets almost always undergo at least one transit in per epoch
In a perfectly isotropic circumbinary population, however, we would expect transits to be more sporadic (see Figure~\ref{fig:setup}).

To go further than this simple intuition, we proceed by investigating the
resulting transit counts distribution as a function of the assumed $\sigma$
distribution. Sampling $\sigma$ across a grid allows us to determine
intermediate cases through an appropriate interpolation, and thus define a
generalised model.

\subsection{Transit Number Distribution at Selected Values of \texorpdfstring{$\sigma$}{sigma} }

As before, we generate and evolve populations of circumbinary planets as
described in Section~\ref{sec:popgen} and apply the filters described in
Section~\ref{sec:filtering}. Furthermore, we do not assume prior knowledge
of the planetary period, and thus sequences filled with many zeros (that
can be explained by longer, aliased periods) are transformed with our
aliasing transformation prescription (see Section~\ref{sec:aliased}).

The results are shown in Table~\ref{tab:scenarioA} - \ref{tab:scenarioD} for
each of the four scenarios described in Section~\ref{sec:results}. Scenarios
B and D are particularly important, since they correspond to the case where
one is simply tallying transits/epoch and is not equipped with special
knowledge of the true period (which we consider to be the typical case
in an automated survey). Scenarios B and D recall are distinct as to whether
broken systems are included (B, yes; D, no) which is dependent upon the
details of the search algorithm in question.

The results reveal how coplanar systems are indeed dominated by
epochs with transits, as expected, but highly misaligned systems are far
more sporadic. For isotropic distributions, we find that of the non-zero
events, over $90$\% produce single transits. With such a high
fraction of single-transit epochs, there is clearly a danger in such
systems being misclassified as planets orbiting single stars - especially
for scenario D where broken systems are discarded in the first place.
Timing variations would of course also be present here but those could be
interpretable as planet-planet interactions. In addition, such systems
will exhibit depth variations for $q<1$. This suggests that it may
be useful to survey the existing catalog of planets ostensibly orbiting
single stars to see if their depth and timing variations might also
be consistent with a misaligned circumbinary scenario.

\subsection{The Relationship between Transit Numbers and \texorpdfstring{$\sigma$}{sigma}}

In practice, one might conceive that inference of $\sigma$ from a population
of circumbinary planets would proceed through a full consideration of the
transit shapes, durations, timing, etc to derive individual system posteriors,
and then these would be fed into a population inference framework, such as
hierarchical Bayesian modelling (e.g. see \citealt{hogg:2010,chen:2017}).
Indeed, population level analyses of the alignment of circumbinary transits
have already been published, such as that of \citet{li:2016} who favour a
nearly coplanar population. However, given the poor sensitivity to misaligned
systems with current methods \citep{martin:2018}, the influence of
selection bias may be driving this.

In what follows, we seek a simplified description for inferring $\sigma$
that depends on the simple and well-defined metric of transit counts per epoch.
Whilst surely more precise inferences of the inclination distribution could
be inferred by leveraging the finer grained information encoded in the
light curve, this metric is attractive for its simplicity, enabling a
relatively straight-forward inference procedure.

To accomplish this, we first need an interpolating function of the grid of
results presented earlier. In what follows, we use the results from
Scenario D, as the likely typical scenario, but the below can be repeated
for any of the other scenarios following the same prescription. Note that in
this scenario, broken sequences are removed and the remaining unbroken
sequences undergo an aliasing transform. This means that there are no
zeros in the final sequences.

As was found in Figure~\ref{fig:brokens}, we find that logistic functions work
well in this problem when the independent variable is $\log \sigma$. This is
because at extreme values of $\sigma$, the distribution saturates to either
perfectly isotropic or perfectly coplanar. Thus, we regressed logistic function
fits to the 4 non-zero columns in the middle panel of Table~\ref{tab:scenarioD}. Additionally, we impose
the constraint that probabilities must be greater than 0, less than 1, and
sum to 1. To impose the ``sum to unity'' constraint, we choose to fit
columns 3, 4, 5 of the middle panel first under the constraint that they sum to less than unity,
then calculate the value of column 2 by hand\footnote{We also tried other
combinations of these but found this to provide the best match to our
simulation results.}.

\begin{figure*}
\includegraphics[width=16.4cm]{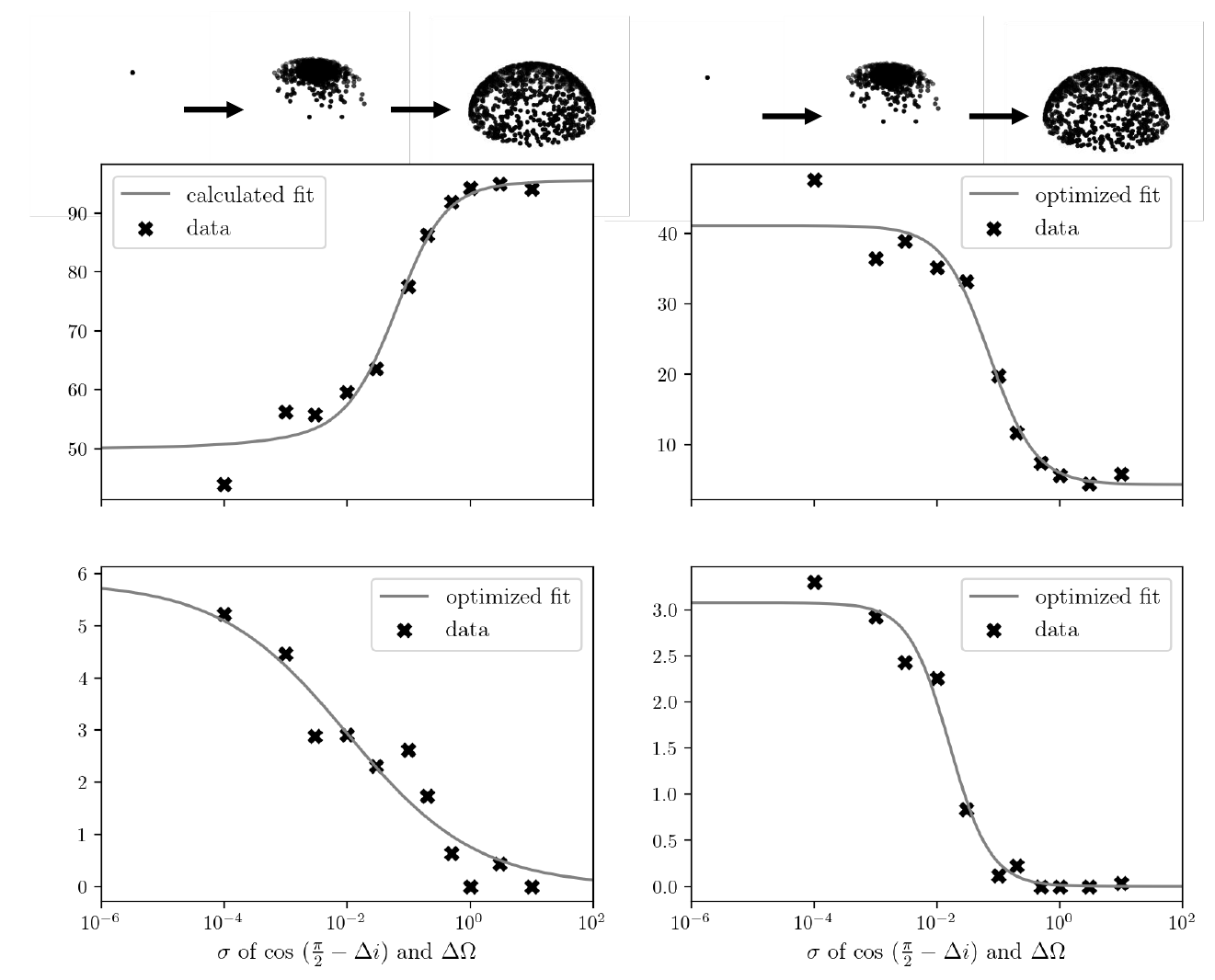}
\caption{
Interpolating functions of the data shown in middle panel of
Table~\ref{tab:scenarioD} (scenario D), using logistic functions. The
one-transit case is defined by deduction of the other three and imposing the
condition that probabilities must sum to 1.
}
\label{fig:multipanel}
\end{figure*}

The resulting fits are shown in Figure~\ref{fig:multipanel}, where
one can see the close agreement achieved between the logistic functions
and the simulation results. Note that even for column 2 where we manually
computed the values instead of directly optimising the fit, the errors are
still quite small, which validates our fitting strategy. 

The logistic function fits are as follows, where $f_j(\sigma)$ is the
fraction of epochs exhibiting $j$ transits:

\begin{align}
f_2(\sigma) &= \frac{0.367}{1+ e^{2.66(\log_{10} \sigma + 1.15)}} + 0.0437,\\
f_3(\sigma) &= \frac{0.0584}{1+ e^{9.54(\log_{10} \sigma + 1.99)}},\\
f_4(\sigma) &= \frac{0.0307}{1+ e^{3.00(\log_{10} \sigma + 1.80)}}
\end{align}

and by definition

\begin{align}
f_1(\sigma) = 1 - f_2(\sigma) - f_3(\sigma) - f_4(\sigma).
\end{align}

Naturally, the above could be easily repeated for the other scenarios,
where we note that one would need to include an $f_0$ function to
account for zero-transit epochs.

\subsection{Constraining the Inclination Distribution}
Consider that we have observed $N$ transit epochs of $M$ distinct transiting
circumbinary planets. The $M$ objects form a population, for example a
group of known circumbinary systems with similar stellar parameters. Let us
assume that each member of the population has a unique $\Obin$ and $\ibin$
value, and that that the ensemble of these values can be approximately
described by the truncated normal distributions posited in
Section~\ref{sec:popgen} and depicted in Figure~\ref{fig:histogram}. In
this way, each object has a unique orbital alignment but the population
is characterised by a single parameter - $\sigma$.

From the sample of $N$ epochs, we can count the number of cases where just
a single transit is observed, $n_1$, or two transits, $n_2$, etc. Defining
$n_4$ as the number of of 4+ transits, we expect $N = n_0 + n_1 + n_2 + n_3
+ n_4$. Note that, in general, $n_0$ is not directly observed, but rather
inferred from the lack of a transit and an inferred ephemeris. However,
in our case of assuming scenario D, it is not possible to have a zero-transit
epoch as discussed earlier so we can discard and define $N = \sum_{k=1}^4 n_k$.
Equipped with the data $\mathcal{D} = \{n_1,n_2,n_3,n_4\}$, we might ask -
what is the allowed range for $\sigma$ that can accommodate these observations?

To make this inference, it is necessary to first define a likelihood
function, $\mathcal{L}$. The likelihood of obtaining the data $\mathcal{D}$,
given a choice of $\sigma$, can be expressed as a multinomial probability mass
function, since the problem is discretised:

\begin{align}
\underbrace{\mathrm{Pr}(\mathcal{D}|\sigma)}_{=\mathcal{L}} =\propto \prod_{k=1}^4 \binom{ \sum_{j=1}^k n_j }{ n_k } f_k^{n_k}.
\end{align}

Since the binomial term in the above is simply a constant for a given set of
data, $\mathcal{D}$, the objective function for inference can be expressed
as

\begin{align}
\log\mathcal{L} &= \sum_{k=1}^4 n_k \log f_k.
\label{eqn:loglike}
\end{align}

\begin{figure}
\includegraphics[width=\columnwidth]{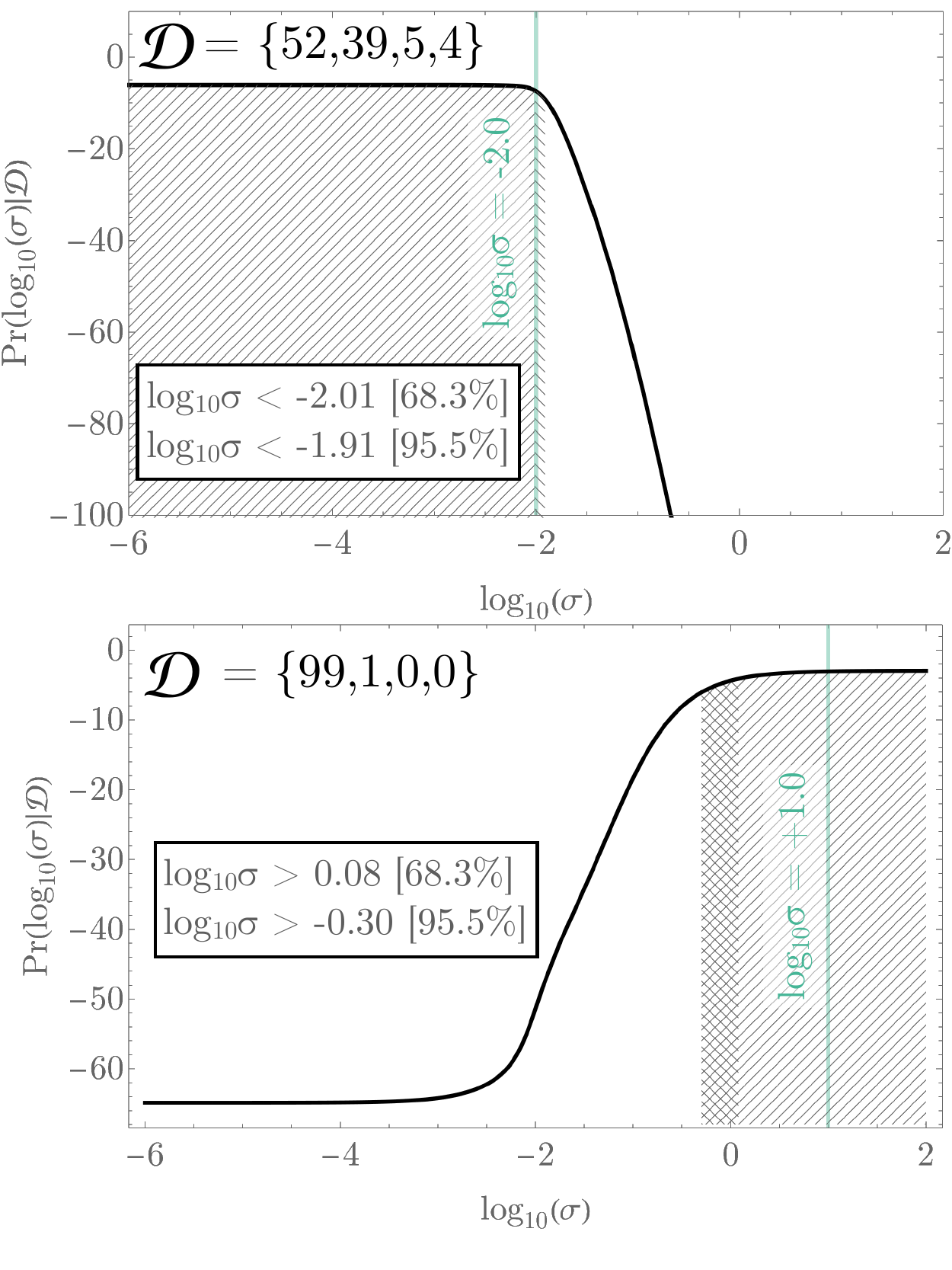}
\caption{
Top: Likelihood of obtaining the transit per epoch counts, represented
by $\mathcal{D}$, as a function of the assumed $\sigma$ value
where the injected truth is $\sigma=0.01$, marked by the green line.
The hatched area denotes the 68.3\% credible interval, and the cross
hatched area marks the 95.5\% credible interval.
Bottom: Same, but for $\sigma=10$ injected truth.
}
\label{fig:inference}
\end{figure}

The likelihood function is typically a monotonic function of $\sigma$ and
thus a central value of $\sigma$ is not well-defined (see
Figure~\ref{fig:inference}). Instead, it is typically more useful to state that
$\sigma$ tends toward the coplanar/isotropic asymptote, and then assign a
statistical upper/lower limit on $\sigma$ parameter, respectively.

To show an example, we set $\sigma = 0.01$ and sampled $N=100$ epochs to
produce the fake data set $\mathcal{D} = \{n_1=52,n_2=39,n_3=5,n_4=4\}$.
Once again, we emphasise that because we are working in scenario D, there are
no zero transit epochs by construction. The likelihood function is shown
in the top panel of Figure~\ref{fig:inference}, plateaus at small $\sigma$
and suggests a nearly coplanar solution. Credible intervals for 68.3\% and
95.5\% are shaded in hatch and cross-hatch respectively, which are
found to be consistent with the injected truth. The exercise was repeated
for $\sigma=10$ and produced another self-consistent recovery (see lower panel of Figure~\ref{fig:inference}), despite the fact
the likelihood function flips over and favours an isotropic solution here.

The inference model here is simplistic and could surely be improved by
leveraging other information within each epoch (e.g. transit durations),
labelling the transits by star and considering alternative population models
beyond a truncated normal. But our simple demonstration establishes that
transit sequences already provide constraining power on the population's
inclination distribution and the problem is tractable.

\section{Discussion}
\label{sec:discussion}

In this work, we have considered the possibility that there exists a population
of misaligned circumbinary transiting planets and explored the observational
consequences. In particular, we have explored the chronological integer sequences
of the number of transits per epoch. Transit flagging is possible through visual
inspection (e.g. \citealt{eisner:2021}) or automated detection, even in the case
of arbitrarily shaped dips \citet{wheeler:2019}, and the number of transits per
epoch would simply be the temporal clusterings of these events. Thus, we expect
that the number of dips occurring in each epoch is a readily observable quantity,
available without significant photodynamical analysis (e.g.
\citealt{doyle:2011}). In this context though, misaligned circumbinaries present
ostensibly foreign sequences when viewed from the familiar perspective of planets
orbiting single stars. We have shown that zero, one, two or three transits per
epoch are not atypical, due to the possibility of missing the stars and the
binary's inner motion. These sequences are so seemingly strange to conventional
search algorithms that they risk being missed altogether \citep{martin:2018}, and thus our work
lays out the expected frequency, patterns and applicability of them.

We have produced synthetic populations of such transit sequences under various
assumptions about the inclination distribution. However, the corresponding
sequences depend upon the details of the detection algorithms that will ultimately
seek them. For example, if sequences include zero-transit events, it may appear
seemingly impossible to draw a quasi-linear ephemeris through the data even
after clustering them into epochs - so-called ``broken'' sequences (e.g.
``1101''). This are the tell-tale signature of a misaligned circumbinary but of
course few algorithms are designed to accommodate the possibility of missing
epochs like this, and thus they may not be ever reported due to detection bias.

We also show how even if the broken sequences can be detected, the naively
inferred period (maximum period which can explain the non-zero epochs) may be wrong
if the sequences are spaced with quasi-regular zero's. We dub these ``aliased
sequences''. For example, the sequence ``10101'' would be naively observed as just
``111'' assuming the period is twice that of reality. That example represents an
unbroken sequence, but the problem is of course exacerbated for unbroken sequences.
This aliasing issue could be resolved using photodynamics, radial velocity or
astrometry analysis but such features are rarely integral part of how transit
algorithms function.

We show that for nearly isotropic distributions of the binary misalignment
angles, zero-transit epochs dominate. With ${\sim}80$\% of epochs being a zero,
our work establishes that any search algorithm aiming for completeness will have
to account for this possibility. Yet more, the occurrence of two or more transits
per epoch, the typical giveaway that one is looking at a transiting circumbinary,
is quite rare - occurring in less than 1\% of epochs. Thus, highly misaligned
circumbinary systems may be hiding in plain sight in our data sets, producing
sporadic single transits of varying depths, durations and irregular spacing.

On the more optimistic side, we have established that these sequences will not
appear as purely random integer chains, and that even two epochs can accurately
predict the third. In principle then, a successful prediction of the next
epoch could establish the system is consistent with a circumbinary. However,
our algorithm is trained on cases where the underlying inclination distribution
is known (e.g. isotropic) and thus in practice applying this to real data would
involve even sifting through a selection of possible inclination distributions,
or modifying as needed. As emphasised throughout, we do not claim this algorithm
is in any way optimal, but is rather used to simply demonstrate the sequences
are non-random in nature (and thus could surely be improved upon).

Finally, we have shown that these sequences encode information about the
inclination distribution, allowing us to in principle test whether a given
set of transit sequences are consistent with an isotropic or coplanar distribution.
That information could of course aid in the prediction issue mentioned earlier.
But, as before, the inference could surely be improved upon by leveraging a
full photodyanamical analysis - the point here is really just to demonstrate that
information is encoded with it and is accessible with even a simple model.

Together, our work highlights the challenges and opportunities that misaligned
transiting circumbinary systems present, taking the observational view that
we often know little more than the timing of transits producing strange
patterns (e.g, \citealt{rappaport:2019}). As transit surveys grow in scope and
size, the need to consider such systems will likely similarly grow.
\section*{Acknowledgements}
We thank the anonymous reviewers for their constructive feedback, which significantly improved this work. This paper made use of data collected by the \kepler\ mission and accessed from the Data Release 25 (DR25) Kepler Input Catalog queried using the Mikulski Archive for Space Telescopes (MAST). Funding for the \kepler\ mission is provided by the NASA Science Mission directorate. We also gratefully
acknowledge the developers of the following software packages
which made this work possible: \rebound\ \citep{rein:2012}, \astropy\ \citep{astropy:2013}, \numpy\ \citep{numpy:2011,numpy:2020}, \scipy\ \citep{scipy:2020}, and \matplotlib\ \citep{matplotlib:2007}.
Simulations in this paper made use of Columbia’s Habanero High Performance Computation Cluster.

We thank the Cool Worlds Lab team members for useful discussions and comments. This work was enabled thanks to supporters of the Cool Worlds Lab, including
Mark Sloan,
Douglas Daughaday,
Andrew Jones,
Elena West,
Tristan Zajonc,
Chuck Wolfred,
Lasse Skov,
Graeme Benson,
Alex de Vaal,
Mark Elliott,
Methven Forbes,
Stephen Lee,
Zachary Danielson,
Chad Souter,
Marcus Gillette,
Tina Jeffcoat,
Jason Rockett,
Scott Hannum,
Tom Donkin,
Andrew Schoen,
Jacob Black,
Reza Ramezankhani,
Steven Marks,
Philip Masterson,
Gary Canterbury,
Nicholas Gebben,
Joseph Alexander \&
Mike Hedlund.

\section*{Data Availability}

The code we used to generate and evolve circumbinary planet populations and the data we obtained is made publicly available at this \href{https://github.com/ziruichen11/misaligned_circumbinary_planets}{URL}. 



\bibliographystyle{mnras}
\bibliography{example} 




%
%


\bsp	
\label{lastpage}
\end{document}